\documentclass[letterpaper]{article}
\usepackage{latexsym,amsfonts,amsmath,amsthm,amssymb}

\begin{document}

\title{A model of quantum-like decision-making with applications to psychology and cognitive science}
\author{Andrei Khrennikov\\
International Center for
Mathematical Modeling \\ in Physics and Cognitive Sciences\\
University of V\"axj\"o, S-35195, Sweden}

\maketitle
\begin{abstract}
We consider the following model of decision-making by cognitive
systems. We present an algorithm -- quantum-like representation
algorithm (QLRA) -- which provides a possibility to represent
probabilistic data of any origin by complex probability
amplitudes. Our conjecture is that cognitive systems developed the
ability to use QLRA. They operate with complex probability
amplitudes, mental wave functions. Since the mathematical
formalism of QM describes as well (under some generalization)
processing of such quantum-like (QL) mental states, the
conventional quantum decision-making scheme can be used by the
brain. We consider a modification of this scheme to describe
decision-making in the presence of two ``incompatible'' mental
variables. Such a QL decision-making can be used in situations
like Prisoners Dilemma (PD) as well as others corresponding to so
called disjunction effect in psychology and cognitive science.
\end{abstract}

\section{Introduction}

Recently a new wave of interest to applications of the
mathematical formalism of QM (especially its QI part) was
generated via interactions of the quantum community with various
research groups working in  artificial intelligence\cite{Bruza},
cognitive science and psychology
\cite{KhrennikovF}--\cite{Franco1}, finances
\cite{PPP}-\cite{Haven} and economy \cite{DAN}, \cite{DAN1} (cf.
with \cite{Whitehead}--\cite{KhrennikovAN}). In particular, an
important project was started in \cite{Jerome1}--\cite{Jerome2},
namely, creation of quantum-like (QL) models for decision-making
by cognitive systems, see also \cite{Franco},\cite{Franco1}. Since
QL-modelling of cognition has always been one of my favorable
domains of research \cite{KhrennikovF}--\cite{Conte}, I was happy
to contribute to this project on decision-making by QL cognitive
systems, see \cite{KHJIP}. In this paper I shall combine the QL
cognitive model \cite{KHJIP}  with Bayesian statistical inference
in the general framework of quantum decision-making, cf. e.g. ,
see e.g. \cite{Hel}, \cite{Holevo}-\cite{Holevo2}, \cite{Mar} (and
references in these works). So, we shall proceed in the same
direction as Busemeyer \cite{Jerome1}-- \cite{Jerome2}, La Mura
\cite{Pr}, \cite{Pr1}  and Franco \cite{Franco}, \cite{Franco1}.

We consider the following model of decision-making by cognitive
systems. We present an algorithm -- quantum-like representation
algorithm (QLRA) -- which provides a possibility to represent
probabilistic data of any origin by complex probability
amplitudes. Our conjecture is that cognitive systems developed the
ability to use QLRA. Thus they operate with complex probability
amplitudes, mental wave functions. Since the mathematical
formalism of QM describes as well (under some generalization, see
appendix -- section 9) processing of such QL mental states, the
conventional quantum decision-making scheme can be used by the
brain. We consider a modification of this scheme to describe
decision-making in the presence of two ``incompatible'' mental
variables. Such a QL decision-making can be used in situations
like Prisoners Dilemma (PD) , see appendix (section 9), as well as
others corresponding to so called disjunction effect in psychology
and cognitive science, see e.g. \cite{G}--\cite{TS}.

We start this paper with a short recollection of the QL
representation of contexts which is based on QLRA, see
\cite{Khrennikov/Supp}, \cite{KhrennikovF1} for detailed
presentation.

\section{Contexts, observables, QL-representation}

\subsection{V\"axj\"o contextual model}

Classical as well as quantum probabilistic models can be obtained
as particular cases of our general contextual model, the
{\it{V\"axj\"o model}}, see \cite{Khrennikov/Supp}.

A physical, biological, social,  mental, genetic, economic, or
financial {\it context} $C$ is  a complex of corresponding
conditions. Contexts are fundamental elements of any contextual
statistical model.\footnote{In principle, the notion of context
can be considered as a generalization of a widely used notion of
{\it preparation procedure}, see e.g. \cite{Bal}, \cite{Bush},
\cite{Holevo}. However, identification of context with preparation
procedure would restrict essentially our theory. In applications
outside physics (e.g., in psychology and cognitive science) we
will consider mental contexts. Such contexts are not simply
preparation procedures. The same can be said about economical,
political and social contexts.  In this book we shall not provide
a deeper formalization of the notion of context. In our model the
notion of context is basic and irreducible.}

Thus construction of any probabilistic model $M$ should be started
with fixing the collection of  contexts of this model. Denote the
collection of contexts by the symbol ${\cal C}$ (so the family of
contexts ${\cal C}$ is determined by the model $M$ under
consideration). Another fundamental element of any contextual
statistical model $M$ is a set of observables ${\cal O}:$ each
observable $a\in {\cal O}$ can be measured under each complex of
conditions $C\in {\cal C}.$ For an observable $a \in {\cal O},$ we
denote the set of its possible values (``spectrum'') by the symbol
$X_a.$ We do not assume that all these observables can  be
measured simultaneously. To simplify considerations, we shall
consider only discrete observables and, moreover, all concrete
investigations will be performed for {\it dichotomous
observables.}

\medskip

{\bf Axiom 1:} {\it For  any observable $a \in {\cal O}$  and its
value $y \in X_a,$ there is defined a context, say $C_y,$
corresponding to the $y$-selection\footnote{See appendix --
section 11 -- for discussion of selection contexts and contextual
forms of the von Neumann projection postulate.}: if we perform a
measurement of the observable $a$ under the complex of physical
conditions $C_y,$ then we obtain the value $a=y$ with probability
1. We assume  that the set of contexts ${\cal C}$ contains
$C_y$-selection contexts for all observables $a\in {\cal O}$ and
$y \in X_a.$}

\medskip

{\bf Axiom 2:} {\it Contextual (conditional) probabilities
$p_C^a(y)\equiv {\bf P}(a=y\vert C)$ are defined  for any context
$C \in {\cal C}$ and any observable $a \in {\it O}.$}

\medskip

Thus, for any context $C \in {\cal C}$ and any observable $a \in
{\it O},$ there is defined the probability to observe the fixed
value $a=y$ under the complex of conditions $C.$ Especially
important role will be played by the ``transition probabilities'':
$ p^{b\vert a}(x\vert y)\equiv {\bf P}(b=x\vert C_y), a, b \in
{\cal O}, y \in X_a, x \in X_b,$ where $C_y$ is the
$[a=y]$-selection context. By axiom 2, for any context $C\in {\cal
C},$ the set of probabilities: $
 \{ {\bf P}(a=y\vert C): a \in {\cal O}\}$ is well defined.
 We complete this probabilistic data for the context $C$  by
transition probabilities. The corresponding collection of data
$D({\cal O}, C)$ consists of contextual probabilities: $ {\bf
P}(a=y\vert C),{\bf P}(b=x\vert C), {\bf P}(b=x\vert C_y), {\bf
P}(a=y\vert C_x),..., $ where $a,b,... \in {\cal O}.$ Finally, we
denote  the family of probabilistic data $D({\cal O}, C)$ for all
contexts $C\in {\cal C}$ by the symbol ${\cal D}({\cal O}, {\cal
C}).$

{\bf Definition 1.} (V\"axj\"o Model) {\it An observational
contextual statistical model of reality is a triple $M =({\cal C},
{\cal O}, {\cal D}({\cal O}, {\cal C})),$ where ${\cal C}$ is a
set of contexts and ${\cal O}$ is a  set of observables which
satisfy to axioms 1,2, and ${\cal D}({\cal O}, {\cal C})$ is
probabilistic data about contexts ${\cal C}$ obtained with the aid
of observables belonging ${\cal O}.$}

We call observables belonging the set ${\cal O}\equiv {\cal O}(M)$
{\it reference of observables.} Inside of a model $M$  observables
belonging  to the set ${\cal O}$ give the only possible references
about a context $C\in {\cal C}.$

{\bf Definition 2.} {\it Let  $a,b \in {\cal O}.$ The observable
$a$  is said to be supplementary\footnote{It might be better to
call such observables complementary, but Bohr's complementarity
was rigidly coupled with mutual exclusivity. Our supplementarity
may be considered as a version of complementarity, but without
mutual exclusivity, see \cite{Khrennikov/Supp}.}  to the
observable $b$ if $ p^{b\vert a}(x\vert y)\not= 0,$ for all $x \in
X_b, y\in X_a.$}

\medskip

\subsection{Law of total probability and its violations}

We recall this law in the simplest case of dichotomous random
variables, $a=y_1,y_2$ and $b=x_1,x_2$, see e.g. \cite{Shir}:
\begin{equation} \label{F} P(b=x)= P(a=y_1) P(b=x\vert a=y_1) +
P(a=y_2) P(b=x\vert a=y_2)
\end{equation}Thus the probability $P(b=x)$ can
be reconstructed on the basis of conditional probabilities $P(b=x
\vert a=y)$ and known or a priori chosen probabilities
$P(a=y).$\footnote{ ``The prior probability to obtain the result
e.g. $b=x_1$ is equal to the prior expected value of the posterior
probability of $b=x_1$ under conditions $a=y_1, y_2.''$} This
formula plays the fundamental role in modern science. Its
consequences are strongly incorporated in modern scientific
reasoning. In
\cite{Khrennikov/book:2004}--\cite{Khrennikov/Supp}it was pointed
out that the quantum formalism induces a modification of this
formula. An additional term appears in the right hand side of
(\ref{F}), so called {\it interference term.}
\begin{equation} \label{FQ} P(b=x)=
P(a=y_1) P(b=x \vert a=y_1) + P(a=y_2) P(b=x\vert a=y_2)
\end{equation}
$$
+ 2\cos\theta \sqrt{P(a=y_1) P(b=x\vert a=y_1)P(a=y_2) P(b=x\vert
a=y_2)}.
$$
The main mathematical consequence of
\cite{Khrennikov/book:2004}--\cite{Khrennikov/Supp}is that any
violation of the formula of total probability (which need not be
coupled to quantum physics) induces its interference
generalization. However, not any violation induces the ordinary
$\cos$-interference. For some contexts violation of (\ref{F})
induces so called hyperbolic interference. But we shall not
consider this type of interference in the present paper.

\section{Quantum-like representation algorithm -- QLRA}

We consider two dichotomous supplementary reference observables
$a$ and $b.$ In \cite{Khrennikov/Supp} we derived the following
formula for interference of contextual probabilities for the
general V\"axj\"o Model:
\begin{equation}
\label{TFR} p_C^b(x) = \sum_y p_C^a(y) p^{b\vert a}(x\vert y) + 2
\lambda_x \; \sqrt{\prod_y p_C^a(y) p^{b\vert a}(x\vert y)},
\end{equation}
where the coefficient of supplementarity (interference):
\begin{equation}
\label{KOL6}  \lambda_x = \frac{p_C^b(x) - \sum_y p_C^a(y)
p^{b\vert a}(x\vert y)}{2\;  \sqrt{\prod_y p_C^a(y) p^{b\vert
a}(x\vert y)}} .
\end{equation}
Contexts such that the interference coefficients $\lambda_x, x\in
X_b,$ are bounded by one are called trigonometric, because in this
case we have the conventional formula of trigonometric
interference:
\begin{equation}
\label{TNCZ} p_C^b(x) = \sum_y p_C^a(y) p^{b\vert a}(x\vert y) + 2
\cos\theta_x \; \sqrt{\prod_y p_C^a(y) p^{b\vert a}(x\vert y)},
\end{equation}
where $ \lambda_x=\cos \theta_x. $ Parameters $\theta_x$ are said
to be $b \vert a$- phases with respect to the context $C.$ We
defined these phases purely on the basis of probabilities. We have
not started with any linear space; in contrast we shall define
geometry from probability.

We denote the collection of all trigonometric contexts by the
symbol ${\cal C}^{\rm{tr}}.$ By using the elementary formula: $
D=A+B+2\sqrt{AB}\cos \theta=\vert \sqrt{A}+e^{i
\theta}\sqrt{B}|^2, $ for real numbers $A, B > 0, \theta\in [0,2
\pi],$ we can represent the probability $p_C^b(x)$ as the square
of the complex amplitude (Born's rule):
\begin{equation}
\label{Born} p_C^b(x)=\vert \psi_C(x) \vert^2 \;.
\end{equation}
Here
\begin{equation}
\label{EX1} \psi(x) \equiv \psi_C(x)= \sqrt{p_C^a(y_1)p^{b\vert
a}(x\vert y_1)} + e^{i \theta_x} \sqrt{p_C^a(y_2)p^{b\vert
a}(x\vert y_2)}, \; x \in X_b.
\end{equation}
The formula (\ref{EX1}) gives the QL representation algorithm --
QLRA. For any trigonometric context $C$ by starting with the
probabilistic data -- $ p_C^b(x), p_C^a(y), p^{b\vert a}(x\vert
y)$ -- QLRA produces the complex amplitude $ \psi_C.$ This
algorithm can be used in any domain of science to create the
QL-representation of probabilistic data (for a special class of
contexts).

We denote the space of functions: $\psi: X_b\to {\bf C}$ by the
symbol $\Phi =\Phi(X_b, {\bf C}).$ Since $X= \{x_1, x_2 \},$ the
$\Phi$ is the two dimensional complex linear space. By using QLRA
 we construct the map $J^{b \vert a}:{\cal C}^{\rm{tr}}
\to \Phi(X, {\bf C})$ which maps contexts (complexes of, e.g.,
physical conditions) into complex amplitudes. The representation
({\ref{Born}}) of probability is nothing other than the famous
{\bf Born rule.} The complex amplitude $\psi_C(x)$ can be called a
{\bf wave function} of the complex of physical conditions
(context) $C$  or a  (pure) {\it state.}  We set
$e_x^b(\cdot)=\delta(x- \cdot)$ -- Dirac delta-functions
concentrated in points $x= x_1, x_2.$ The Born's rule for complex
amplitudes (\ref{Born}) can be rewritten in the following form:
$\label{BH} p_C^b(x)=\vert \langle \psi_C, e_x^b \rangle \vert^2,$
where the scalar product in the space $\Phi(X_b, C)$ is defined by
the standard formula: $\langle \phi, \psi \rangle = \sum_{x\in
X_b} \phi(x)\bar \psi(x).$ The system of functions
$\{e_x^b\}_{x\in X_b}$ is an orthonormal basis in the Hilbert
space $H=(\Phi, \langle \cdot, \cdot \rangle).$

Let $X_b \subset {\bf R}.$ By using the Hilbert space
representation  of the Born's rule  we obtain  the Hilbert space
representation of the expectation of the observable $b$: $E(b
\vert C)= \sum_{x\in X_b} x\vert\psi_C(x)\vert^2= \sum_{x\in X_b}
x \langle \psi_C, e_x^b\rangle \overline{\langle\psi_C,
e_x^b\rangle}= \langle \hat b \psi_C, \psi_C\rangle,$ where the
(self-adjoint) operator $\hat b: H \to H$ is determined by its
eigenvectors: $\hat b e_x^b=x e^b_x, x\in X_b.$ This is the
multiplication operator in the space of complex functions
$\Phi(X_b,{\bf C}):$ $ \hat{b} \psi(x) = x \psi(x).$ It is natural
to represent the $b$-observable (in the Hilbert space model)  by
the operator $\hat b.$

We would like to have Born's rule not only for the $b$-variable,
but also for the $a$-variable: $p_C^a(y)=\vert \langle \psi, e_y^a
\rangle\vert^2 \;, y \in  X_a.$

How can we define the basis $\{e_y^a\}$ corresponding to the
$a$-observable? Such a basis can be found starting with
interference of probabilities. We set $u_j^a=\sqrt{p_C^a(y_j)},
p_{ij}=p^{b \vert a}(x_j \vert y_i), u_{ij}=\sqrt{p_{ij}},
\theta_j=\theta_C(x_j).$ We have:
\begin{equation}
\label{0} \psi=u_1^a e_{y_1}^a + u_2^a e_{y_2}^a,
\end{equation}
where
\begin{equation}
\label{Bas} e_{y_1}^a= (u_{11}, \; \; u_{12}) ,\; \; e_{y_2}^a=
(e^{i \theta_1} u_{21}, \; \; e^{i \theta_2} u_{22})
\end{equation}

Suppose now that the matrix of transition probabilities $P^{b\vert
a}$ is doubly stochastic.\footnote{It is a square matrix of
nonnegative real numbers, each of whose rows and columns sums to
1. Thus, a doubly stochastic matrix is both left stochastic and
right stochastic.} Under this condition  the system
$\{e_{y_i}^a\}$ is an orthonormal basis iff the probabilistic
phases satisfy the constraint: $ \theta_2 - \theta_1= \pi \;
\rm{mod} \; 2 \pi.$ In this case the $a$-observable is also
represented by  a self-adjoint  operator $\hat{a}$ which is
diagonal with eigenvalues $y_1,y_2$ in the basis $\{e_{y_i}^a\}.$
The conditional average of the observable
 $a$ coincides with the quantum Hilbert space average:
$ E(a \vert C)=\sum_{y \in X_a} y p_C^a(y) = \langle \hat{a}
\psi_C, \psi_C \rangle. $

In the general case (when $P^{b\vert a}$ need not be doubly
stochastic) the $a$-observable is represented as a generalized
quantum observable (non self-adjoint operator), see appendix
(section 10). We remark that statistical data obtained in
cognitive psychology in experimental tests of disjunction effect
produce non doubly stochastic matrices of transition probabilities
\cite{TS}, \cite{ST}.

\section{QL Decision-making scheme}

As we have seen, if for some context $C,$  probability
distributions for supplementary observables $a$ and $b$ are known,
then the complex probability amplitude $\psi_C$  representing $C$
can be reconstructed by using QLRA.  This was the problem of
representation of probabilistic data by complex probability
amplitude, see section 3. My conjecture is that the brain
developed the ability for such a QL representation of
probabilistic data, see \cite{KHJIP} for details. In such aq
QL-model the brain uses complex probability amplitudes for
decision-making.

We consider the following situation. A (mental) context $C$ is
given. The brain must take decision about the $b$-attribute, given
by e.g. $b=x_1, x_2,$  -- so to choose between  $b=x_1$ and
$b=x_2.$ The crucial point is that it is assumed that another
attribute, say $a (=y_1,y_2),$ which is supplementary to $b,$ is
involved in the process of decision-making. Since variables $a$
and $b$ are supplementary (under the context $C),$ interference
angles $\theta=(\theta_{x_1},\theta_{x_2})$ should be considered,
see (\ref{EX1}). In the PD , see appendix (section 10), this
$a$-attribute is related to actions of another prisoner. In the
gambling experiment it is simply the (classical) random generator
producing wins and losses. The latter example shows that
``quantumness'' (qualitatively encoded by the interference angles)
is not a feature of $a$ (in fact neither of $b),$ but it appears
via interrelation of $a, b$ and the context $C.$ Our scheme of  QL
decision-making is based on the assumptions that there are given
(created by the brain of the basis of previous experience):

\medskip

a) transition probabilities $p^{b \vert a}(x \vert y);$

b) the probability distribution of the $a: \; p_C^a(y);$

c) the probability distribution of the phase angles
$\theta=(\theta_{x_1}, \theta_{x_2}) : \; p_C(\theta).$

\medskip

Thus all these distributions are given a priori. One should not
always identify prior probabilities with ``subjective
probabilities.'' The previous frequency experience plays an
important role in determination of these probability
distributions, cf.  \cite{Cox}.

The brain uses the formula of total probability with the
interference term to find the $b$-probabilities. Under the
assumption that the interference angle is $\theta_x,$ it produces
the probabilities
$$
p_C^b(x\vert \theta)= \sum_y p_C^a(y) p^{b \vert a}(x \vert y) + 2
\cos \theta_x\;  \sqrt{p_C^a(y_1) p^{b \vert a}(x \vert
y_1)p_C^a(y_2) p^{b \vert a}(x \vert y_2)}.
$$
The crucial point of the decision-making scheme is their
interpretation:

\medskip

{\it For each $x,$ $p_C^b(x \vert \theta)$ is the probability that
under the condition that the $b\vert a$-interference angle is
$\theta_x$ (for the context $C)$ the decision $b=x$ is ``right'',
i.e., it would produce some form of reward.}

\medskip

By  the (classical) Bayes' formula the brain finds the joint
probability distribution:
\begin{equation} \label{DDD1} p_C(x, \theta)=
p_C(\theta)\Big(\sum_y p_C^a(y) p^{b \vert a}(x \vert y) + 2 \cos
\theta_x\; \sqrt{p_C^a(y_1) p^{b \vert a}(x \vert y_1)p_C^a(y_2)
p^{b \vert a}(x \vert y_2)}\Big)
\end{equation}
and finally the total $b$-probabilities \begin{equation}
\label{ST1} \bar{p}_C^b(x)= \int d \theta \; p_C(\theta) \;  p_C^b
(x \vert \theta).
\end{equation}
As the  extension of the interpretation of conditional
probabilities, the probability $\bar{p}_C^b(x)$ is considered as
the probability that the decision $b=x$ is right.

\medskip

In the present decision-making scheme the brain makes the
$b=x_1$-decision if $\bar{p}_C^b(x_1)$ is larger than
$\bar{p}_C^b(x_2)$ an vice versa, cf. \cite{Holevo}, p. 54. The
qualitative meaning of ``larger'' is determined depending on the
cognitive system and may be the context $C.$

\medskip

We should also mention another QL decision-making scheme.
Comparing of the probabilities $\bar{p}_C^b(x_1)$ and
$\bar{p}_C^b(x_2)$ is an additional act of mental processing. It
needs special neuronal and time recourses. The processing might be
especially complicated when these probabilities do not differ
essentially. In such a situation a QL cognitive system might
choose the regime of ``automatic probabilistic decision-making",
namely, by just using a (classical) random generator producing
decisions $x_1$ and $x_2$ with the probabilities
$\bar{p}_C^b(x_1)$ and $\bar{p}_C^b(x_2)$

\medskip

{\bf Remark 1.} (Comparing with classical probability) We remark
that a cognitive system $\tau_{CL}$ which uses the classical
probabilistic processing of information can apply the conventional
formula of total probability (\ref{F}) to predict the
$b$-probabilities on the basis transition probabilities  $p^{b
\vert a}(x \vert y)$ and $a$-probabilities  $p_C^a(y).$ Thus one
can consider the proposed QL-scheme as simply introduction of an
additional -- interference -- parameter $\theta$ and modification
of the formula of total probability. The main source of such a
modification of the conventional statistical considerations is the
impossibility to combine the context $C$ with the selection
contexts $C_{y_j}$ and hence to get the probabilities $P(b=x \vert
C C_{y_j}),$ cf. with the resolution of ``Simpson's paradox'' in
\cite{Lindley}. As we have seen, a QL cognitive system $\tau_{QL}$
cannot proceed in the same way. The formula of total probability
with the interference term contains not only the transition
probabilities and the $a$-probabilities, but also phases and the
latter are unknown. Thus even by choosing e.g. prior probabilities
$p_C^a(y)$ (under the condition that the transition probabilities
were obtained via the frequency experience), the $\tau_{QL}$ could
not predict $b$-probabilities.

\medskip

By using QLRA the cognitive system $\tau_{QL}$ can construct for
each $\theta= (\theta_{x_1},\theta_{x_2})$ the complex probability
amplitude $\psi_{C,\theta}(x).$ Then the $b$-probabilities can be
represented by using the Born's rule:
\begin{equation}
\label{ST1Y} \bar{p}_C^b(x)= \int d \theta \; p_C(\theta) \; \vert
\psi_{C,\theta}(x) \vert^2.
\end{equation}

\section{Bayesian updating of state distribution}

Thus by our model the brain of $\tau_{QL}$ proceeds by using the
mixture of classical and quantum of probabilities. The whole
Bayesian scheme is purely classical, ``quantumness'' appears in
(\ref{ST1Y}) only via Born's rule.

However, as always, there arises the problem of the choice of
prior probability distributions. Since the transition
probabilities and the $a$-probabilities are present even in the
classical Baeysian framework, only the phase distribution
$p_C(\theta)$ makes a new (QL) contribution. A QL cognitive system
$\tau_{QL}$ should learn itself to choose $p_C(\theta)$ on the
basis of the previous experience of the $b\vert a$ decision-making
under the context $C.$ Such a learning can be performed via the
(conventional) Bayesian updating procedure.

By combining Bayes' and Born's formulas, we get:
\begin{equation} \label{DDD}
p_C(\theta \vert x)= \frac{p_C(x, \theta)}{\bar{p}_C^b(x)}=
\frac{p_C(\theta) \vert \psi_{C,\theta}(x) \vert^2}{ \int d \theta
p_C(\theta) \vert \psi_{C,\theta}(x) \vert^2}.
\end{equation}
By following the Bayesian scheme $\tau_{QL}$ would like to
maximize the probability $p_C(\theta \vert x),$ i.e., to construct
a map $m: X_b \to \Theta, m(x) = \theta_{\rm{max}}(x),$ see
\cite{Cox}. Since the denominator in (\ref{DDD}) does not depend
on $\theta,$ this problem is reduced to maximization of  the joint
probability density $p_C(x, \theta).$

Suppose now that under the context $C$ the $\tau_{QL}$ made the
decision $b=x$ and this decision was successful (so the
$\tau_{QL}$ got some form of reward). Then the $\tau_{QL}$ would
update the distribution $p_C(\theta)$ by maximizing $p_C(x,
\theta).$ To simplify considerations and to extract the main QL
factor, we assume that the transition probabilities as well as the
$a$-probabilities are fixed. So,  optimization is considered only
with respect to the interference angles $\theta.$

In the case of the doubly stochastic matrix of transition
probabilities $\theta_{x_1}= \theta_{x_2} + \pi$ and hence we can
consider the one dimensional phase parameter $\theta.$

\medskip

{\bf Example 1.} (Discrete distribution of phases) Some context
$C$ is chosen. Suppose that the  transition probabilities  as well
as  the $a$-probabilities are equal to 1/2.  Here the formula of
total probability with the interference term gives:
$$
p_C(x_1 \vert \theta)= \cos^2 \theta/2; p_C(x_2, \theta)=  \sin^2
\theta/2.
$$
We remark that these probabilities coincide with polarization (or
spin 1/2) probabilities obtained in QM, see e.g. \cite{Holevo}. It
should be emphasized that this is really a simple coincidence of
mathematical formulas. In opposite to e.g. \cite{Mar}, we do not
consider physical quantum systems. We now consider the simplest
nontrivial case of the parametric set consisting of two points,
e.g. $\Theta=\{\theta_1=\pi/2,\theta_2=\pi\}.$ So, this cognitive
system reduced (on the basis of some information) phases under the
context $C$ to two possible angles. Hence,
$\bar{p}_C(x_1)=\frac{1}{2} ( \cos^2  \pi/4 + \cos^2 \pi/2)
=\frac{1}{4}, \bar{p}_C(x_2)=\frac{1}{2} (\sin^2 \pi/4 + \sin^2
\pi/2)=\frac{3}{4}.$ Thus under the assumption that all phases in
$\Theta$ are equally possible, this cognitive system $\tau_{QL}$
gets that $\bar{p}_C(x_2)$ is essentially larger than
$\bar{p}_C(x_1).$ Hence, $\tau_{QL}$  makes the decision $b=x_2.$
If the result of this decision was positive (i.e. some form of
reward was obtained), $\tau_{QL}$ would like to update the state
distribution. Since $p_C(x_2, \pi/2)=\frac{1}{4}$ and $p_C(x_2,
\pi/2)= \frac{1}{2},$ the cognitive system will put (in future
decision-making) more weight to $\theta_2= \pi,$  e.g. the updated
distribution could be $p_C(\pi/2)=\frac{1}{3},
p_C(\pi)=\frac{2}{3}.$

\medskip

{\bf Example 2.} (Continuous distribution of phases)  Suppose that
all transition probabilities are equal. Let us consider the
uniform distribution of phases on $\Theta=[0, 2\pi): d
p_C(\theta)= \frac{1}{2\pi} d \theta.$ Here $p_C(x_1, \theta)=
\frac{1}{2\pi} \cos^2 \theta/2; p_C(x_2, \theta)= \frac{1}{2\pi}
\sin^2 \theta/2.$ Hence, $\bar{p}_C(x_1)= \bar{p}_C(x_2)=1/2.$
Thus the definite decision could not be done.

{\bf Example 3.} Suppose that all transition probabilities are
equal. Let us consider the uniform distribution of phases on
$\Theta=[0, \pi/2): d p_C(\theta)= \frac{2}{\pi} d \theta.$ Here
$p_C(x_1, \theta)= \frac{2}{\pi} \cos^2 \theta/2; p_C(x_2,
\theta)= \frac{2}{\pi} \sin^2 \theta/2.$ Hence,
$\bar{p}_C(x_1)=\frac{1}{\pi}+ \frac{1}{2},
\bar{p}_C(x_2)=\frac{1}{2}- \frac{1}{\pi}.$ Thus the $b=x_1$
decision is preferred. For this decision the maximum is approached
for $\theta=0.$ Therefore this cognitive system would update
$p_C(\theta)$ by concentrating it at the point $\theta=0.$

\section{Mixed state representation}

We remark that the former Bayesian considerations can be
mathematically represented by using mixed quantum states. Let us
consider the density matrix: $$\rho_C \equiv \int_\Theta d \theta
\; p(\theta)  \;\rho_{C, \theta} .
$$
$$
\rho_{C, \theta}\equiv \psi_{C, \theta} \otimes  \psi_{C, \theta}
$$
 We obtain the
representation:
 \begin{equation} \label{ST2} \bar{p}_C^b(x)=\rm{Tr}\;
\rho_C \;\pi^b_x,
\end{equation}
where $\pi^b_x$ is the orthogonal projector
corresponding to the eigenvalue $b=x.$ Thus quantity
\begin{equation}
\label{ST3}
\frac{\bar{p}_C^b(x_1)}{\bar{p}_C^b(x_2)}=\frac{\rm{Tr} \;\rho_C\;
\pi^b_{x_1}}{\rm{Tr}\; \rho_C \;\pi^b_{x_2}}
\end{equation}
is used in the QL decision-making.

\section{Comparing with standard quantum decision-making theory}

In this section we would like to compare our approach with
standard quantum decision-making theory, see e.g. \cite{Hel},
\cite{Holevo}-\cite{Holevo2}, \cite{Mar} (and references in these
works).:

a). Interpretation. The crucial difference is that our formalism
is not about really quantum physical systems, but about QL
systems. Thus we need not quantum sources of randomness, e.g.
electrons or photons, to perform our QL decision making. Moreover,
the essence of QL behavior is not consideration of a special class
of systems, but of a special class of contexts or to be more
precise: interrelation between contexts and observables.

b). Scheme of the decision making. We consider a specific scheme
(motivated by PD, see appendix (section 9)) involving {\it two
supplementary (``incompatible'')  observables} $a$ and $b.$
Moreover, in general one of them, namely, $a$  is a generalized
quantum observable, see section 10.

c). Mathematics. We consider a specific parametrization of a prior
quantum state, namely, by the interference angle $\theta.$

d). Application. We apply our model to modelling of brain's
functioning as a macroscopic QL system or to be more precise: a
macroscopic system performing specific interconnections between
contexts and observables (inducing nontrivial interference).

\section{Bayes risk}

As usual in quantum decision-making, we consider Bayes risk
corresponding to the deviation function $W_\theta(x),$ see
\cite{Holevo}, p. 46:
\begin{equation}
\label{LT} {\cal R}_C^b\equiv \int_\Theta d p(\theta)\; \sum_x
W_\theta(x) p_C^b(x\vert \theta)= \int_\Theta d p(\theta)\; \sum_x
W_\theta(x) \; \vert \psi_{C,\theta}(x) \vert^2=
\end{equation}
$$
\int_\Theta d p(\theta)\; \sum_x W_\theta(x) \; \; \rm{Tr}
\rho_{C, \theta} \; \pi_x^b.
$$
Typically in quantum decision theory the problem of finding of
Bayes decision rule is considered, e.g. \cite{Holevo}, p. 46--50.
However, we are not interested in this problem, since the
decision-making operator $\hat{b}$ is considered as
given.\footnote{Of course, it could also be modified in the
process of brain's functioning, but we do not consider this
problem in the present paper.}

In our model the brain is interested to minimize Bayes risk for
the fixed observable $b$ via variation of the prior distribution
of interference phases.

We come back to Example 1. Now we do not fix the distribution of
phases on $\Theta=\{\theta_1=\pi/2,\theta_2=\pi\}.$ Here
$p=p(\theta_1)$ and $1-p=p(\theta_2)$ are parameters of the model.
Suppose that the deviation function
$W_{\theta_j}(x_i)=\delta_{ij}.$ Thus Bayes risk is $ {\cal
R}_C^b= p\; p_C^b(x_1\vert \theta_1) + (1-p)\; p_C^b(x_2\vert
\theta_2)= p \cos^2 \theta_1/2+ (1-p) \sin^2 \theta_2/2=
p/2+(1-p)= 1-p/2.$ Thus Bayes risk is minimal for $p=1.$ Hence,
the brain would modify the prior (mixed)  mental state into the
(pure) mental state $\psi_{C,\pi/2}.$

\section{Prisoner's Dilemma}

In game theory, PD is a type of non-zero-sum game in which two
players can cooperate with or defect (i.e. betray) the other
player. In this game, as in all game theory, the only concern of
each individual player (prisoner) is maximizing his/her own
payoff, without any concern for the other player's payoff. In the
classic form of this game, cooperating is strictly dominated by
defecting, so that the only possible equilibrium for the game is
for all players to defect. In simpler terms, no matter what the
other player does, one player will always gain a greater payoff by
playing defect. Since in any situation playing defect is more
beneficial than cooperating, all rational players will play
defect.

The classical PD is as follows: Two suspects, $A$ and $B,$ are
arrested by the police. The police have insufficient evidence for
a conviction, and, having separated both prisoners, visit each of
them to offer the same deal: if one testifies for the prosecution
against the other and the other remains silent,
 the betrayer goes free and the silent accomplice receives the full 10-year sentence.
If both stay silent, both prisoners are sentenced to only six
months in jail for a minor charge. If each betrays the other, each
receives a two-year sentence. Each prisoner must make the choice
of whether to betray the other or to remain silent. However,
neither prisoner knows for sure what choice the other prisoner
will make. So this dilemma poses the question: How should the
prisoners act? The dilemma arises when one assumes that both
prisoners only care about minimizing their own jail terms. Each
prisoner has two options: to cooperate with his accomplice and
stay quiet, or to defect from their implied pact and betray his
accomplice in return for a lighter sentence. The outcome of each
choice depends on the choice of the accomplice, but each prisoner
must choose without knowing what his accomplice has chosen to do.
In deciding what to do in strategic situations, it is normally
important to predict what others will do. {\it This is not the
case here.} If you knew the other prisoner would stay silent, your
best move is to betray as you then walk free instead of receiving
the minor sentence. If you knew the other prisoner would betray,
your best move is still to betray, as you receive a lesser
sentence than by silence. Betraying is a dominant strategy. The
other prisoner reasons similarly, and therefore also chooses to
betray. Yet by both defecting they get a lower payoff than they
would get by staying silent. So rational, self-interested play
results in each prisoner being worse off than if they had stayed
silent, see e.g. wikipedia -- ``Prisoner's dilemma.'' The
following mental contexts are involved in PD:

Context $C$ representing the situation such that a player has no
idea about planned action of another player. Context $C_{+}^a$
representing the situation such that the $B$-player supposes that
$A$ will cooperate and context $C_{-}^a$ -- $A$ will compete. We
can also consider similar contexts $C_{\pm}^b.$ We define
dichotomous observables $a$ and $b$ corresponding to {\it actions}
of players $A$ and $B:$ $a=+$ if $A$ chooses to cooperate and
$a=-$  if $A$ chooses to compete, $b$ is defined in the same way.

A priori the law of total probability might be violated for PD,
since the $B$-player is not able to combine contexts. If those
contexts were represented by subsets of a so called space of
``elementary events'' as it is done in classical probability
theory (based on Kolmogorov (1933) measure-theoretic axiomatics),
the $B$-player would be able to consider the conjunction of the
contexts $C$ and  e.g. $C_{+}^a$ and to operate in the context $C
\wedge C_{+}^a$ (which would be represented by the set $C \cap
C_{+}^a).$ But the very situation of PD is such that one could not
expect that contexts $C$ and  $C_{\pm}^a$ might be peacefully
combined. If the $B$-player obtains information about the planned
action of the $A$-player (or even if he just decides that $A$ will
play in the definite way, e.g. the context $C_{+}^a$ will be
realized), then the context $C$ is simply destroyed. It could not
be combined with $C_{+}^a.$

We can introduce the following contextual probabilities:
$p_C^b(\pm)\equiv P(b=\pm \vert C)$ -- probabilities for actions
of $B$ under the complex of mental conditions $C.$ $p^{b \vert
a}(\pm,+) \equiv P(b=\pm \vert C_+^a)$ and $p^{b \vert
a}(\pm,-)\equiv P(b=\pm \vert C_-^a)$ -- probabilities for actions
of $B$ under the complexes of mental conditions $C_+^a$ and
$C_-^a,$ respectively, $p_C^a(\pm)\equiv P(a=\pm \vert C)$ --
prior probabilities which $B$ assigns  for actions of $A$ under
the complex of mental conditions $C.$

\section{Appendix: Generalization of the  QM  formalism}

Let us consider a finite dimensional Hilbert space $H.$ Let ${\cal
E}=\{e_j\}_{j=1}^n$ be an orthonormal basis:
\begin{equation} \label{BBBB}
\psi=\sum_j c_j e_j, c_j=c_j(\psi) \in {\bf C}.
\end{equation}
 Each ${\cal E}$
generates a class of (conventional) quantum observables,
self-adjoint operators, see \cite{VN}, \cite{D}:
\begin{equation} \label{SA} \hat{a} \psi=\sum_j y_j c_j(\psi) e_j,
\end{equation}
where $X_a=\{y_1, ..., y_n\}, y_j \in {\bf R},y_j\not= y_i $ is
the range of values of $a$ (so we start with consideration of
observables with nondegenerate spectra).

Let now ${\cal E}=\{e_j\}_{j=1}^n$ be an arbitrary basis (thus in
general $\langle e_j, e_i \rangle \not=0, i \not= j)$ consisting
of normalized vectors, i.e., $\langle e_j, e_j
\rangle=1.$\footnote{We remark that QLRA, see section 3, produces
the $a$-basis with normalized vectors, $\Vert e_y^a \Vert^2=1.$ It
is a consequence of stochasticity of an arbitrary matrix of
transition probabilities (which was used by QLRA to produce the
$a$-basis). Thus we consider now a purely linear algebraic
version of this situation.}

 We generalize the Dirac-von Neumann formalism by considering
observables (\ref{SA}) for an arbitrary ${\cal E}.$ We also
consider an arbitrary nonzero vector of $H$ as a pure quantum
state. We postulate (by generalizing Born's postulate):
\begin{equation} \label{SA1} P_\psi(a=y_j)= \frac{\vert
c_j(\psi)\vert^2}{\sum_j\vert c_j(\psi)\vert^2},
\end{equation}
where the coefficients $c_j(\psi)$ are given by the expansion
(\ref{BBBB}).

 If ${\cal E}$ is an orthonormal basis, then
$c_j(\psi)=\langle \psi, e_j \rangle, \sum_j\vert
c_j(\psi)\vert^2= \Vert \psi \Vert^2$ and for a normalized vector
$\psi,$ we obtain the ordinary Born's rule.

Our generalization of the Dirac-von Neumann formalism is also very
close to another well known (and very popular in QI)
generalization of the class of quantum observables, namely, to the
formalism of POVM, \cite{Bush}, \cite{Holevo}. To proceed in this
way, we introduce projectors on the basis vectors: $\pi_j
\psi=c_j(\psi) e_j.$ We remark that $\pi_j^2= \pi_j,$ but in
general $\pi_j^*\not= \pi_j.$ We have: $\vert c_j(\psi)\vert^2=
\langle \pi_j \psi, \pi_j \psi\rangle= \langle M_j \psi,
\psi\rangle,$ where $M_j= \pi_j^* \pi_j.$ We remark that each
$M_j$ is self-adjoint and, moreover, positively defined. We also
set $M=\sum_j M_j.$ Then our generalization of Born's rule can be
written as: \begin{equation} \label{SA2} P_\psi(a=y_j)=
\frac{\langle M_j \psi, \psi\rangle }{\langle M \psi,
\psi\rangle}= \frac{\rm{Tr}\; \rho_\psi M_j}{\rm{Tr}\; \rho_\psi
M},
\end{equation}
where $\rho_\psi= \psi\otimes\psi.$ We remark that, for an
arbitrary nonzero $\psi,$ the operator $\rho_\psi\geq 0.$

Now we generalize the conventional notion of the density operator,
by considering any nonzero $\rho \geq 0$ as a generalized density
operator (we recall that at the moment we consider a
finite-dimensional space). The corresponding generalization of
Born's postulate has the following form:
\begin{equation} \label{SA3} P_\psi(a=y_j)= \frac{\rm{Tr}\; \rho\;
M_j}{\rm{Tr}\; \rho \;M}.
\end{equation}
The only difference from the POVM formalism is that the operator
$M\not=I$ (the unit operator).

We remark that $\langle M\psi, \psi \rangle= \sum_j \vert
c_j(\psi) \vert^2 \not= 0, \psi \not=0.$ Thus (we are in the
finite dimensional case) the inverse operator $M^{-1}$ is well
defined.

We now proceed with our formalization and consider an arbitrary
(separable) Hilbert space $H.$

\medskip

{\bf Definition 10.1.} {\it A generalized quantum state is
represented by an arbitrary trace class nonnegative (nonzero)
operator $\rho: \rho \geq 0, 0 < \rm{Tr} \rho < \infty.$}

\medskip

{\bf Definition 10.2.} {\it A generalized quantum observable is
represented by an arbitrary (so in general non normalized)
positive operator valued measure $E$ on a measurable space
$(X,{\cal F})$ such that $E(X)>0.$}

\medskip

Thus, for a generalized quantum observable $E,$   we have:

1). $E(B)\geq 0,$ for any set $B\in {\cal F},$ and $E(X)>0;$

2). $E(\cup_{j=1}^n B_j)=\sum_{j=1}^n E(B_j)$ for all disjoint
sequences $\{B_j\}$ in ${\cal F}.$

\medskip

{\bf Generalized Born's rule:} Let $\rho$ and $E$ be generalized
quantum state and observable, respectively. Then the probability
to find the result $x$ of the $E$-measurement  in a measurable set
$B$ (for an ensemble represented by $\rho$) is given by
\begin{equation}
\label{QST} P_\rho(x \in B)= \frac{\rm{Tr} \rho \;E(B)}{\rm{Tr}
\rho \;E(X)}.
\end{equation}

We remark that $\rm{Tr} \rho \;E(X)>0.$ To prove this, we consider
the spectral expansion of  the trace class operator $\rho= \sum_j
q_j \psi_j\otimes\psi_j.$ Here at least one $q_j>0.$ Then $\rm{Tr}
\rho \;E(X)=\sum_j q_j \langle E(X)\psi_j, \psi_j\rangle>0.$

We now come back to the model considered at the beginning of this
section: a finite-dimensional space. We would like to model in the
abstract linear algebra framework the situation considered in
section 3. We consider two observables, one is a conventional
self-adjoint operator $\hat{b}$ and another is a generalized
observable $\hat{a}.$ Thus the $b$-basis ${\cal E}^b=\{e_j^b\}$ is
orthonormal, but the $a$-basis ${\cal E}^a=\{e_j^a\}$ need not
(but we emphasize that even the latter one is normalized). Any
vector $e_j^a$ is a conventional (pure) quantum state. Thus by the
rules of the conventional QM we can find ``transition
probabilities'': $p^{b \vert a}(x_i \vert y_j)= P_{e_j^a}(b=x_i)=
\vert \langle e_j^a, e_i^b \rangle \vert^2.$ Since ${\cal E}^b$ is
orthonormal, we have: $\sum_i p^{b \vert a}(x_i\vert y_j))=\sum_i
\vert \langle e_j^a, e_i^b\rangle \vert^2 = \Vert e_j^a \Vert^2=
1.$ The matrix of $b \vert a$-transition probabilities $P^{b \vert
a}$ is stochastic (as it should be). However, if ${\cal E}^a$ is
not orthonormal, then $P^{b \vert a}$ is not doubly stochastic.

On the other hand, we can expand each $e_i^b$ with respect to
${\cal E}^a: e_i^b= \sum_j c_j(e_i^b) e_j^a.$ By our generalized
Born's rule: $p^{a\vert b}(y_j\vert x_i)= P_{e_i^b}(a=y_j)= \vert
c_j(e_i^b)\vert^2/ \sum_j \vert c_j(e_i^b)\vert^2.$ We have:
$\sum_j p^{a\vert b}(y_j\vert x_i)=1.$  Thus even the matrix of
transition probabilities $P^{a \vert b}$ is stochastic.

Finally, we remark that all previous considerations are valid even
in the case when both observables are generalized.

\section{Appendix: Von Neumann postulate in cognitive science and psychology}

In  general the transition probabilities can depend on the
cognitive context $C$ which was chosen for the first
(unconditional) measurement:
$$
p^{b \vert a}(x \vert y) =p^{b \vert a}_{C}(x\vert y).
$$
But in some cases dependence of the transition probabilities $p^{b
\vert a}_{C}(x\vert y)$ on $C$ could be reducible. In the
experimental situation these probabilities (frequencies) are found
in the following way. First cognitive systems interact with a
context $C.$ In this way  an ensemble $S_C$ of cognitive systems
representing the context $C$ is created. Then cognitive systems
belonging to the ensemble $S_C$ interact with the
selection-context $C_y$ which is determined by the value $y$ of
the mental observable $a.$ For example, students belonging to a
group $S_C$ (which was trained under a complex of mental or social
conditions $C)$ should answer to the question $a.$ If this
question is so disturbing for a student $\omega$ that he would
totally forget about the previous $C$-training, then the
transition probabilities do not depend on $C: p^{b \vert a}(x\vert
y).$ Since we are interested only in probabilities, such an
individual blocking can be generalized to ``statistical blocking''
-- dependence on $C$ after sequential $ab$-measurement should be
statistically negligible: the number of persons who still use the
original $C$-context (e.g. training)  to reply to the $b$-question
(following the $a$-question) is negligibly small comparing with
the total number of persons in a sample $S_C$ representing $C.$

We remark that this is the case in conventional quantum theory.
Here for incompatible (noncomutative) observables (with
nondegenerate spectra) the transition probabilities $p^{b \vert
a}(x\vert y)=\vert (e_x^b, e_y^a)\vert^2$ do not depend  on the
original context $C,$ i.e.,  a context preceding the $a=y$
selection (by the QM-terminology: ``on the original wave function
$\psi$'').

In quantum theory any $a=y$ selection destroys the memory on the
preceding physical context $C.$  For example, suppose that we
prepare electrons with a wave function $\psi$ (which provides
symbolic symbolic representation of a context $C,$ so $\psi=
\psi_C).$ We measure spin's projection on some direction $b$ and
then on another direction $a.$ The transition probability does not
depend on $\psi$ (i.e., on $C).$

This is our contextual interpretation of the von Neumann
projection postulate  \cite{VN}.

We do not know the general situation for cognitive
systems.\footnote{ It might be that the von Neumann projection
postulate can be violated by cognitive systems. In such a case we
would not be able to construct the conventional quantum
representation of contexts by complex probability amplitudes.}
Our conjecture is that\footnote{We recall that we consider only
observables with nondegenerate spectra.}:

\medskip

{\bf Postulate.} (``von Neumann postulate for mental observable'')
{\it For any pair $a, b$ of supplementary mental observables the
transition probability $p^{b \vert a}(x\vert y)$ is completely
determined by the preceding preparation -- context $C_y$
corresponding to the $[a=y]$-selection.}

\medskip

We remark that by Axiom 1
$$
p^{b\vert b}( x \vert x)=1.
$$
Thus if ``a system was prepared in the state $e_x^b$,'' then
measurement of $a$ would definitely give the value $b=x.$

To proceed in our contextual framework, we could be satisfied even
by a weaker form of this postulate -- we recall that QLRA works by
using only two ``reference observables.''

\medskip

{\bf Postulate.} (``Weak von Neumann postulate for mental
observable'') {\it There exist supplementary  mental observables
$a, b$ such that the transition probability $p^{b \vert a}(x\vert
y)$ is completely determined by the preceding preparation --
context $C_y$ corresponding to the $[a=y]$-selection.}

\end{document}